\documentclass[iop]{emulateapj}
\slugcomment{{\sc Accepted to ApJL:} March 5, 2016}
\usepackage{appendix,natbib}
\usepackage{subfigure}
\usepackage[backref,colorlinks=true,citecolor=blue,linkcolor=blue,urlcolor=blue]{hyperref}
\usepackage[all]{hypcap}

\newcommand{\halpha}{H\ensuremath{\alpha}}
\newcommand{\hbeta}{H\ensuremath{\beta}}

\newcommand{\um}{$\mu$m}
\def\msun{{\rm M_\odot}}

\begin{document}

\title{The {MOSDEF} Survey: The strong agreement between {\halpha} and UV-to-FIR star formation rates for $\lowercase{z}\sim2$ star-forming galaxies\altaffilmark{*}}

\author{\sc Irene Shivaei\altaffilmark{1,7}, Mariska Kriek\altaffilmark{2}, Naveen A. Reddy\altaffilmark{1}, Alice E. Shapley\altaffilmark{3}, Guillermo Barro\altaffilmark{2}, Charlie Conroy\altaffilmark{4}, Alison L. Coil\altaffilmark{5}, William R. Freeman\altaffilmark{1}, Bahram Mobasher\altaffilmark{1}, Brian Siana\altaffilmark{1}, Ryan Sanders\altaffilmark{3}, Sedona H. Price\altaffilmark{2}, Mojegan Azadi\altaffilmark{5}, Imad Pasha\altaffilmark{2}, Hanae Inami\altaffilmark{6}}

\altaffiltext{1}{Department of Physics \& Astronomy, University of California, Riverside, CA 92521, USA}
\altaffiltext{2}{Astronomy Department, University of California, Berkeley, CA 94720, USA}
\altaffiltext{3}{Department of Physics \& Astronomy, University of California, Los Angeles, CA 90095, USA}
\altaffiltext{4}{Department of Astronomy, Harvard University, Cambridge, MA, USA}
\altaffiltext{5}{Center for Astrophysics and Space Sciences, University of California, San Diego, La Jolla, CA 92093, USA}
\altaffiltext{6}{National Optical Astronomy Observatory, 950 N. Cherry Avenue, Tucson, AZ 85719, USA}
\altaffiltext{7}{NSF Graduate Research Fellow}
\altaffiltext{*}{Based on observations made with the W.M. Keck Observatory, which is operated as a scientific partnership among the California Institute of Technology, the University of California, and the National Aeronautics and Space Administration.}

\begin{abstract}

We present the first direct comparison between Balmer line and panchromatic SED-based SFRs for $z\sim2$ galaxies. For this comparison we used 17 star-forming galaxies selected from the MOSFIRE Deep Evolution Field (MOSDEF) survey, with $3\sigma$ detections for {\halpha} and at least two IR bands ({\em Spitzer}/MIPS 24\,{\um} and {\em Herschel}/PACS 100 and 160\,{\um}, and in some cases {\em Herschel}/SPIRE 250, 350, and 500\,{\um}). The galaxies have total IR ($8-1000\,\mu$m) luminosities of $\sim10^{11.4}-10^{12.4}\,\textrm{L}_\odot$ and star-formation rates (SFRs) of $\sim30-250\,\textrm{M}_\odot\,\mathrm{yr^{-1}}$. We fit the UV-to-far-IR SEDs with flexible stellar population synthesis (FSPS) models -- which include both stellar and dust emission -- and compare the inferred SFRs with the SFR({\halpha},{\hbeta}) values corrected for dust attenuation using Balmer decrements. The two SFRs agree with a scatter of 0.17 dex. Our results imply that the Balmer decrement accurately predicts the obscuration of the nebular lines and can be used to robustly calculate SFRs for star-forming galaxies at $z\sim2$ with SFRs up to $\sim200\,\msun\,{\rm{yr^{-1}}}$. We also use our data to assess SFR indicators based on modeling the UV-to-mid-IR SEDs or by adding SFR(UV) and SFR(IR), for which the latter is based on the mid-IR only or on the full IR SED. All these SFRs show a poorer agreement with SFR({\halpha},{\hbeta}) and in some cases large systematic biases are observed. Finally, we show that the SFR and dust attenuation derived from the UV-to-near-IR SED alone are unbiased when assuming a delayed exponentially declining star-formation history.

\end{abstract}
\keywords{dust, extinction --- galaxies: general --- galaxies: high-redshift --- galaxies: star formation --- infrared: galaxies}
\maketitle

\section{Introduction}

Star-formation rates (SFRs) are among the most fundamental measurements for constraining the physics of galaxy formation and evolution. The past decade has seen a multitude of studies that trace SFRs out to high redshift \citep[][ and references therein]{madau14}, and examined their correlation with other galaxy properties, such as stellar masses \citep[e.g.,][]{salim07,reddy12b,shivaei15b}.
The {\em Spitzer Space Telescope} and {\em Herschel Space Observatory} opened a new window into measuring bolometric SFRs by allowing us to directly correct the widely-used SFRs(UV) for dust-obscured star formation \citep[e.g.,][]{barro11,whitaker14b,debarros15}. However, despite their common use, the UV+IR SFRs have several disadvantages: first, due to the limited spatial resolution and sensitivity of {\em Spitzer} and {\em Herschel}, most of the individual distant galaxies are not detectable in IR images, and second, the empirical templates used to convert the mid-IR (restframe 8\,{\um}) fluxes to total IR luminosities result in systematic biases \citep{elbaz11,reddy12a,utomo14}.

Hydrogen Balmer emission lines are considered to be another gold standard for tracing robust SFRs in local galaxies \citep[][among many others]{moustakas06,hao11,kennicutt12}. Until recently, high-redshift studies \citep[e.g.,][]{muzzin10,ibar13,shivaei15a} only detected {\halpha} with no (or very limited) {\hbeta} measurements, which are required for accurate dust corrections. However, with the new generation of multi-object near-IR (NIR) spectrographs, both {\halpha} and {\hbeta} are now detectable for normal star-forming galaxies at $z\sim2$ \citep{kashino13,steidel14,kriek15}.

While dust-corrected SFRs({\halpha}) using Balmer decrements (hereafter, SFRs({\halpha},{\hbeta})) are commonly used at low redshift, we still need to investigate how well Balmer lines trace SFR at high redshifts, as galaxies were more star-forming at $z\sim2$, the peak of the SFR density evolution \citep{hopkins06}. In particular, it has been argued that Balmer lines may miss optically thick star-forming regions at these high redshifts \citep[e.g.,][]{rodighiero14}. In order to investigate this possible bias, one needs to compare the SFRs({\halpha},{\hbeta}) with independently measured UV-to-far-IR (UV-to-FIR) SFRs for star-forming galaxies at $z\sim2$. Fortunately, such dusty systems are among the luminous tail of the galaxy distribution within reach of {\em Herschel}. 

In this paper, we use the unique dataset of the MOSFIRE Deep Evolution Field (MOSDEF) survey \citep{kriek15} in combination with {\em Herschel} and {\em Spitzer} data, to investigate how well SFRs({\halpha},{\hbeta}) agree with UV-to-FIR SFRs at $z\sim2$. For the latter, we use the Flexible Stellar Population Synthesis (FSPS) models \citep{conroy09}, which utilize energy balance to fit the stellar and dust emission simultaneously \citep{berta13,dacunha15}. Throughout this paper, we assumed a \citet{chabrier03} initial mass function (IMF), and H$_0=70\,{\rm km\,s^{-1}\,Mpc^{-1}},\Omega_{\Lambda}=0.7,\Omega_m=0.3$.

\begin{figure}[tbp]
	\subfigure{
		\includegraphics[trim=1cm .2cm 0cm 0cm,width=.52\textwidth]{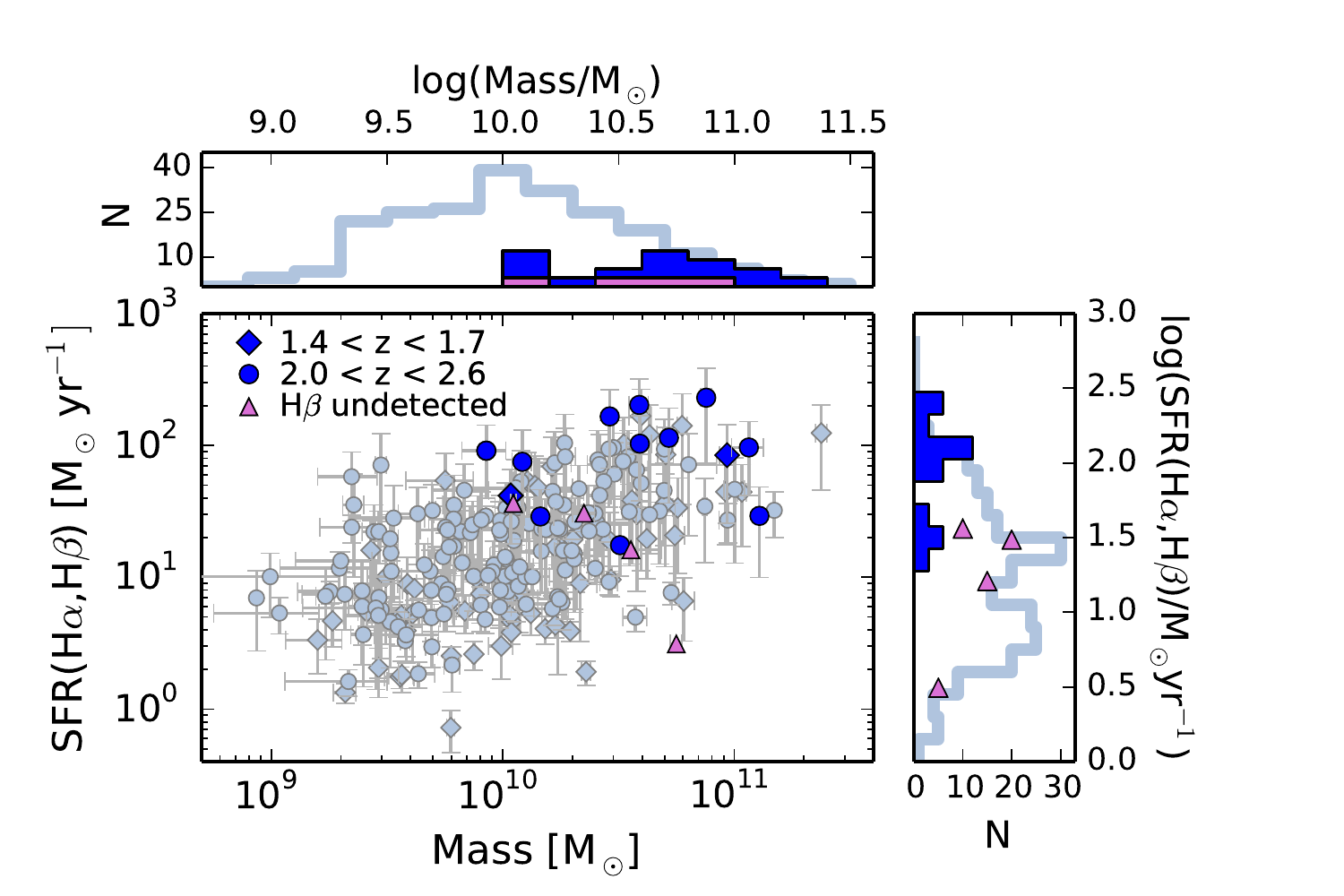}}
	\quad
	\subfigure{
		\includegraphics[trim=1cm .2cm 0cm .7cm,width=.52\textwidth]{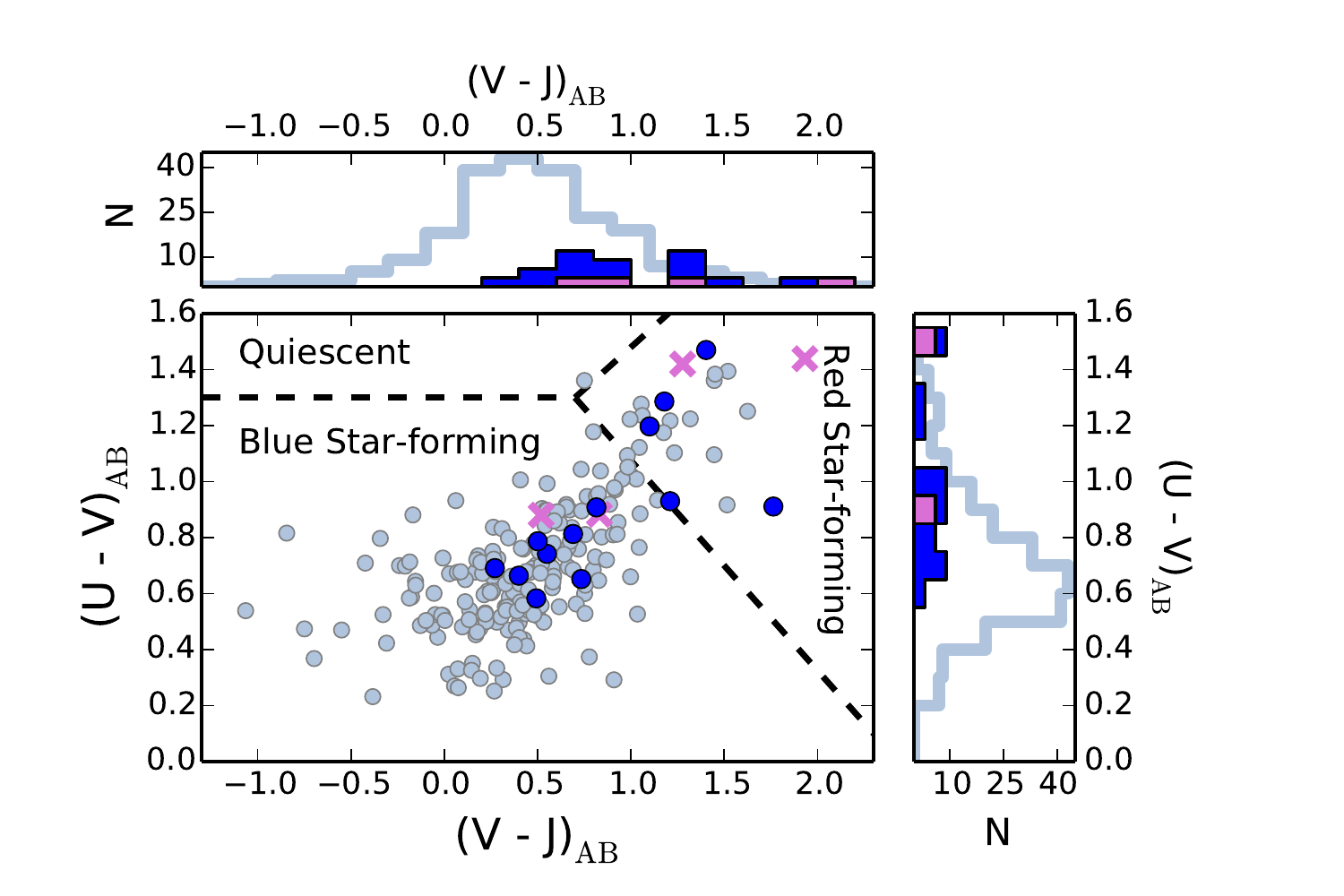}	}
		\caption{SFR({\halpha},{\hbeta}) vs. $M_*$ (top) and rest-frame $UVJ$ diagram (bottom) for the MOSDEF spectroscopic sample (light blue) and the 13 candidates in this study (dark blue). The four undetected-{\hbeta} galaxies are the purple symbols with a $3\sigma$ lower-limit on their SFRs({\halpha},{\hbeta}). 
		For displaying purposes the number counts of the dark blue histograms are multiplied by 3.}
	\label{fig:ms}
\end{figure}

\section{Sample and Measurements}
\label{sample}

\begin{figure*}[tp]
	\centering
		\includegraphics[width=1\textwidth]{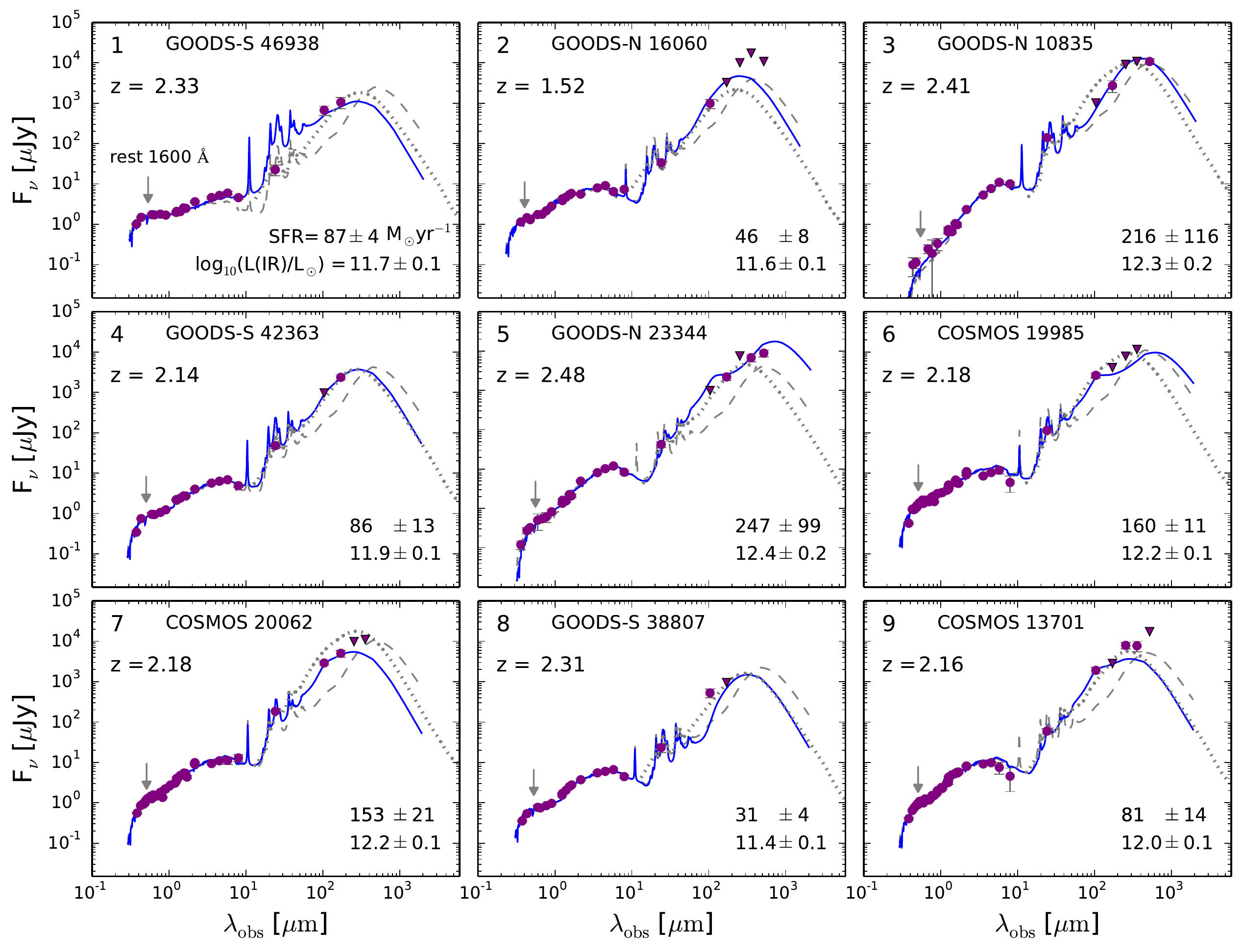}
		\caption{The best-fit panchromatic FSPS SED models (solid curves) to the observed photometry (purple symbols) of 9 of our {\halpha}+{\hbeta}-detected galaxies. The rest-frame optical emission lines are excluded from both the photometry and the models.
		The triangles show $3\sigma$ upper limits on the fluxes; in the fitting process, the actual photometry with the corresponding errors were used.
		The dashed and dotted lines represent the FSPS SED fits up to 24\,{\um} and the \citet{ce01} best-fit templates to the 24, 100, and 160\,{\um} data, respectively. The ID numbers are from the 3DHST v4.1 photometric catalog. The galaxies are shown in order of increasing mass (continued in Figure~\ref{fig:seds2}).
	}
	\label{fig:seds1}
\end{figure*}

\begin{figure*}[tp]
	\centering
		\includegraphics[width=1\textwidth]{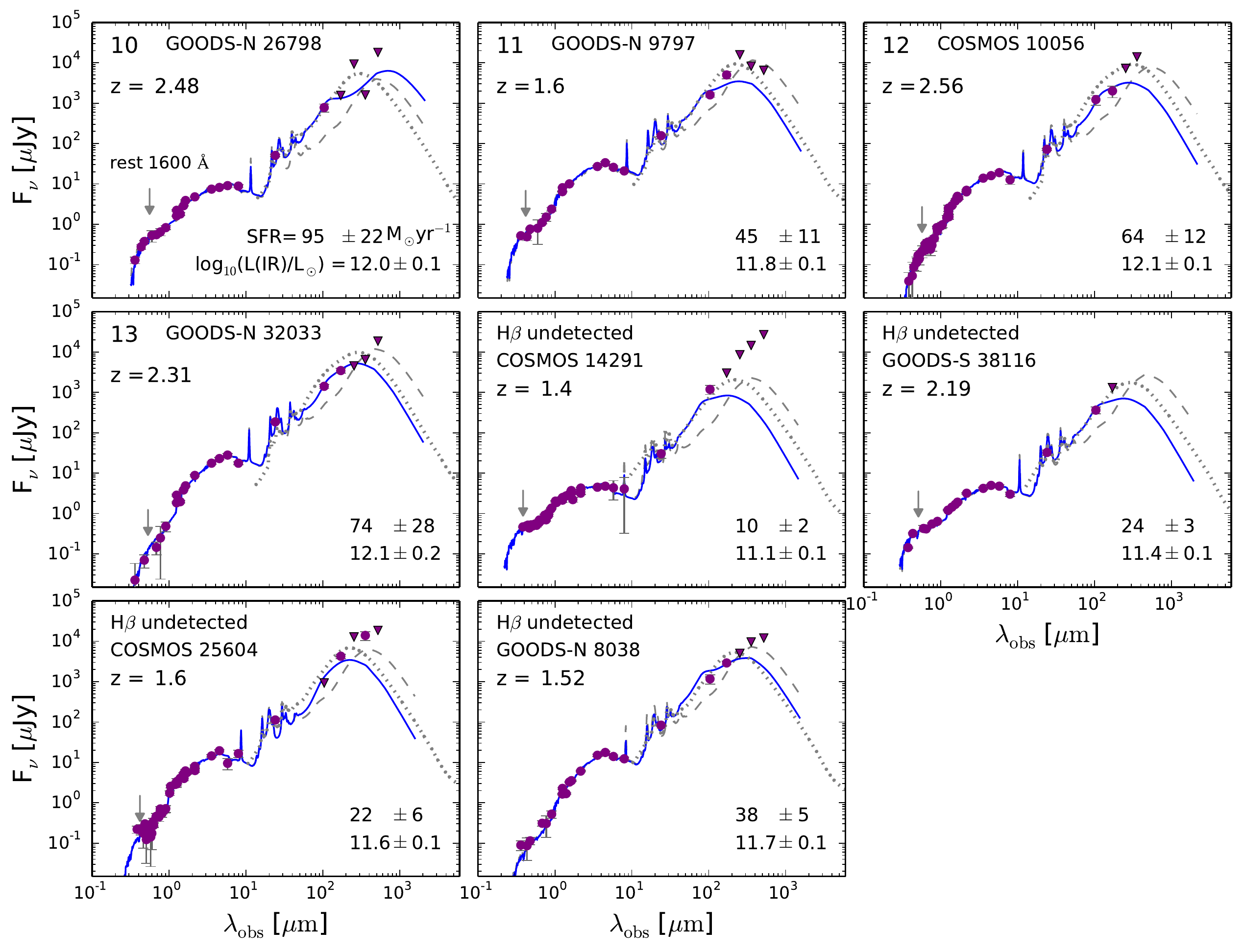}
		\caption{The best-fit SED models of the rest of our {\halpha}+{\hbeta}-detected sample and the 4 {\hbeta}-undetected galaxies. See the caption of Figure~\ref{fig:seds1}. 
	}
	\label{fig:seds2}
\end{figure*}

The MOSDEF survey is a multi-year project that uses the MOSFIRE spectrometer on the Keck~I telescope to study the stellar and gaseous content of $\sim1500$ galaxies and AGNs at $1.37\leq z\leq 3.80$. 
The survey covers the CANDELS fields, for which extensive deep multi-wavelength data are available \citep{grogin11,koekemoer11,momcheva15}. For details of the survey strategy, observations, data reduction, and sample characteristics see \citet{kriek15}. The current study is based on data from the first two years of the MOSDEF survey.

We selected a sample of MOSDEF galaxies based on their Balmer emission and IR luminosities. The {\halpha} and {\hbeta} luminosities were estimated by fitting Gaussian functions to the line profiles. Flux uncertainties were derived using Monte Carlo simulations. The spectra were flux calibrated by comparing the spectrum of a slit star with the total photometric flux \citep{skelton14} in the same filter. To account for the fact that galaxies are resolved, an additional slit-loss correction was applied using the {\em HST} profiles for each galaxy \citep[see][]{kriek15}. The uncertainties on the total slit-loss corrected fluxes were 16\% and 20\% for {\halpha} and {\hbeta}, respectively \citep{reddy15}. {\halpha} and {\hbeta} fluxes were corrected for underlying Balmer absorption as determined from UV-to-NIR SED modeling \citep{reddy15}. The {\halpha} luminosity was converted to SFR using the \citet{kennicutt98} relation and was corrected for dust attenuation using the Balmer decrement ({\halpha}/{\hbeta}), assuming the \citet{cardelli89} Galactic extinction curve. For those galaxies with undetected {\hbeta}, we used $3\sigma$ upper-limits on the {\hbeta} fluxes that translate to lower-limits on {\halpha}/{\hbeta} and thus lower-limits on dust-corrected SFR({\halpha},{\hbeta}).

For the IR measurements we used data from {\em Spitzer}/MIPS and {\em Herschel}/PACS and SPIRE in the COSMOS (PI: M.~Dickinson), GOODS-N, and GOODS-S fields \citep{oliver10,elbaz11,magnelli13}.
We measured IR fluxes simultaneously using scaled point-spread functions (PSFs) \citep{reddy10,shivaei15a}, which enable us to more robustly recover fluxes of confused objects. In brief, for each image we used a prior list of objects from higher resolution images to determine the location of all sources in the field. For the MIPS images we used $3\sigma$ IRAC sources; for the PACS images we used $3\sigma$ MIPS sources; and for the SPIRE images we used the PACS sources detected at $3\sigma$ (for the 250 and 350\,{\um} bands) and $5\sigma$ (for the 500\,{\um} band). In order to perform the photometry, a subimage centered on each target was constructed and the PSF was fitted simultaneously to all the objects in the subimage (including the target) and one random background position, which was chosen to be at least 1 FWHM away from any source. We repeated this procedure for 20 random background positions. The standard deviation of the background fluxes was adopted as the flux uncertainty. 

The accuracy of our photometry was tested by simulating 500 sources with a large range of fluxes that were added to the original images. The simulated fluxes were recovered successfully with a given scatter that, as expected, decreased with increasing flux. Based on this scatter, we estimated a correction factor to be applied to the background-estimated errors for each individual source. These corrections ranged from a factor of $1.3-2.0$, depending on the filter.

We selected galaxies whose IR photometry was not significantly affected by nearby sources and had robust detections (signal-to-noise ratio (S/N) $>3\sigma$) in {\halpha}, MIPS 24\,{\um}, and at least one PACS filter (100 or 160\,{\um}).
The final sample consists of 13 galaxies with both {\halpha} and {\hbeta} detections and 4 {\hbeta}-undetected galaxies. Only three of these galaxies were detected in the SPIRE filters. We did not include objects that were classified as AGNs based on their X-ray emission, IRAC colors, or optical emission lines (with the criterion of [N{\sc ii}]/{\halpha}$>0.5$, \citealt{coil15}).
As seen in Figure~\ref{fig:ms}, most of our galaxies lie above the median SFR-$M_*$ relation by \citet{shivaei15b} with $M_*\sim10^{10}-10^{11}\,\msun$. 

\section{Panchromatic SED Modeling}
\label{sec:fullsed}

In this section we derive total SFRs by comparing the panchromatic SEDs - composed of the UV-to-IRAC photometry from 3D-HST \citep{skelton14,momcheva15} combined with IR photometry (previous section) - with stellar population models.
We generated composite stellar populations using the FSPS code \citep[][v2.5]{conroy09,conroy10}, assuming a Chabrier IMF and solar metallicity. For the star-formation history (SFH) we assumed a delayed-tau model of the form ${\rm SFR(}t{\rm )}\propto te^{-t/\tau}$, in which $t$ is the age of the galaxy and $\tau$ is the e-folding time. $\tau$ was allowed to vary from $10^2-10^4$\,Myr in steps of 0.2\,dex. We set a 50\,Myr lower limit on age corresponding to the typical dynamical timescale at $z\sim2$ \citep{reddy15}. We adopted a \citet{calzetti00} attenuation curve, in which the dust optical depth at 5500\,\AA~was allowed to vary from 0.0 to 4.0 in steps of 0.1. The dust emission in the FSPS code follows the \citet{draine07} (DL07) models. In these dust models, the radiation field strength, $P(U)dU$, is described by a delta function at $U_{\rm min}$ (the radiation intensity to which the dust in the diffuse ISM is exposed) and a power-law component for the radiation from dust heated by more intense starlight (OB associations). The parameter $\gamma$ represents the fraction of dust that is heated by the intense starlight. As was suggested in DL07, we allowed $U_{\rm min}$ to vary between 0.1 to 25.0 in 10 logarithmic steps, the $\gamma$ grid was set to $[0.0,0.01,0.02,0.04,0.06,0.10,0.15]$, and we explored PAH fractions of $0,1,2,3,5,7,10$\%.

We shifted the models to the spectroscopic redshift of each galaxy, applied the \citet{madau95} prescription to correct the SEDs for the intergalactic medium absorption, and then, adopting a $\chi^2$ minimization method (using the original photometric errors), fit the models to the UV-to-FIR photometry. The optical photometry was corrected for contamination by bright emission lines ({\halpha}, [O{\sc iii}], [O{\sc ii}], etc.) as measured from the MOSDEF spectra \citep{reddy15}. We perturbed the photometry according to the measurement uncertainties for 100 realizations and the dispersion in each best-fit parameter was adopted as the associated uncertainty.

The best-fit models of the {\halpha}+{\hbeta}-detected candidates are displayed in Figures~\ref{fig:seds1} and \ref{fig:seds2}. 
We would have found similar SFRs if instead we modeled the IR emission by a single blackbody SED.
In order to assure that our best-fit models are not skewed by the smaller 24\,{\um} errors compared to the PACS errors, we tested the SED fitting with equal weighting for the 24, 100, and 160\,{\um} fluxes. For all but two galaxies (galaxies 1 and 13), the corresponding SFRs and IR luminosities were within $1\sigma$ uncertainty of the original SFR and L(IR) values.

\begin{figure*}[tbp]
	\centering
		\includegraphics[height=.9\textwidth]{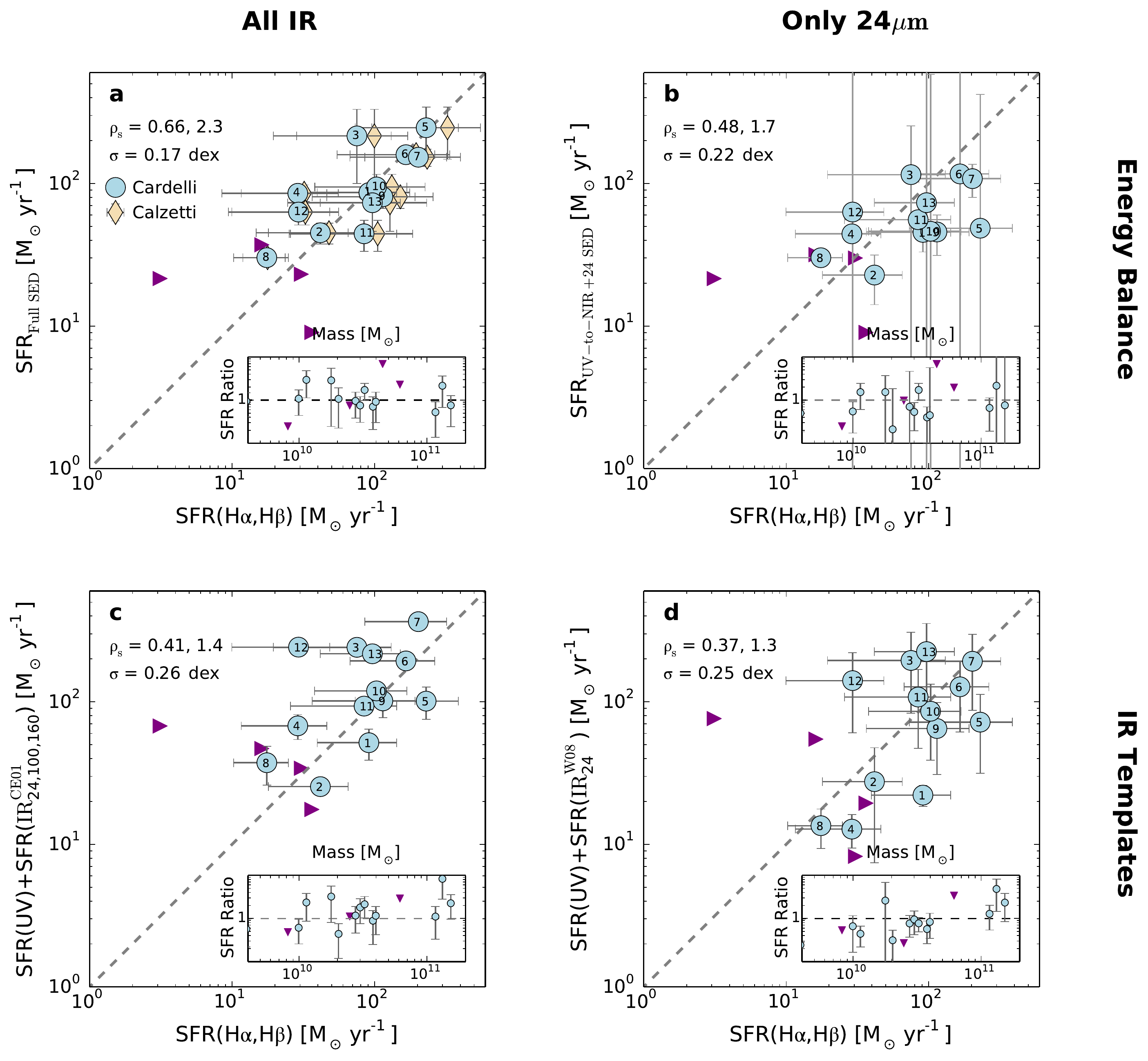}
		\caption{SFR comparisons for the 13 {\halpha}+{\hbeta}-detected galaxies (circles) and the 4 {\hbeta}-undetected galaxies (triangles). In all plots, the horizontal axis is the SFR({\halpha},{\hbeta}), assuming the Cardelli curve, and the dashed lines denote one-to-one relationships. {\b In panel a, diamonds indicate SFR({\halpha},{\hbeta}) assuming the Calzetti curve}. The vertical axis is the SFR derived from modeling the UV-to-FIR photometry (a), the UV-to-24\,{\um} photometry (b), and SFR(UV)+SFR(IR), in which SFR(IR) is derived from $24-160\,\mu$m (c) or 24\,{\um} only photometry (d), using the CE01 and W08 templates, respectively. 
		Each subplot shows the ratio of the SFR in the main plot's vertical axis to SFR({\halpha},{\hbeta}) as a function of $M_*$. The masses are inferred from the best-fit SEDs. The Spearman coefficient ($\rho_{\rm s}$), its significance, and scatter about the unity line ($\sigma$) are listed in the plots. The galaxies' numbers correspond to the SEDs in Figures~\ref{fig:seds1},\ref{fig:seds2}.}
	\label{fig:sfrs}
\end{figure*}

\section{SFR Comparison}
\label{sec:sfrs}

In this section we investigate how the SFRs derived from different diagnostics compare with each other. Fitting the stellar and dust SEDs simultaneously using energy balance is thought to give a robust estimate of the total SFR. Figure~\ref{fig:sfrs}(a) shows that SFRs({\halpha},{\hbeta}) are in good agreement with the full SED-inferred SFRs (SFR$_{\rm{full\,SED}}$). The tightness of the correlation between $\log$(SFR({\halpha},{\hbeta})) and $\log$(SFR$_{\rm{full\,SED}}$)\footnote{The $\log$(SFR({\halpha},{\hbeta})) and $\log$(SFR$_{\rm{full\,SED}}$) relation has 0.17\,dex scatter about the unity line and Spearman $\rho_{\rm s}=0.66$ with $2.3\sigma$ from a null correlation.}, its consistency with the unity line, and the absence of any systematic correlation as a function of $M_*$ (see the inset plot~\ref{fig:sfrs}(a)), imply that the dust-corrected L({\halpha},{\hbeta}) is an unbiased tracer of the total SFR for $z\sim2$ star-forming galaxies with SFRs up to $\sim200\,\msun\,{\rm yr}^{-1}$.

There is a systematic uncertainty associated with our choice of the attenuation curve. Assuming the Calzetti curve instead of the Cardelli curve to correct the observed {\halpha} predicts higher SFRs by an average factor of 1.2 (as high as 1.4 for the galaxy with the highest SFR; diamonds in Figure~\ref{fig:sfrs}\,a). The larger dust correction is expected as the Calzetti curve has a higher normalization at 6563\,\AA. We adopted the Cardelli curve as it was derived based on the line-of-sight measurements of HII regions, and therefore preferable for the extinction correction of nebular lines.

Given that the majority of the $z\sim2$ galaxies are not detected with {\em Herschel}, it is important to know how well we can recover the total SFR with only 24\,{\um} data. We fit the FSPS models to the UV-to-24\,{\um} data and set the dust emission parameters to the default values (PAH fraction = 3.5\%, $\gamma=0.01$, and $U_{\rm min}=1.0$). The result is shown in Figure~\ref{fig:sfrs}(b). There is a larger (0.22\,dex) scatter between SFR$_{\rm{UV-24}}$ and SFR({\halpha},{\hbeta}), and SFR$_{\rm{UV-24}}$ underestimates SFR({\halpha},{\hbeta}) for the highest star-forming galaxies in our sample.

In Figure~\ref{fig:sfrs}, panels c and d, we estimated the bolometric SFR by adding unobscured SFR(UV) to obscured SFR(IR).
The SFR(UV) is derived from luminosity at 1600\,\AA~by fitting a power-law function of the form $f_{\lambda}\propto\lambda^{\beta}$ to the photometry at rest wavelengths $1268-2580$\,\AA, and using the \citet{kennicutt98} conversion between L(UV) and SFR. In order to estimate the total L(IR), we fit the locally calibrated \citet{ce01} templates (hereafter, CE01) to the 24, 100, and 160\,{\um} fluxes in Figure~\ref{fig:sfrs}(c), and in Figure~\ref{fig:sfrs}(d) we converted the 24\,{\um} to total L(IR) using the luminosity-independent conversions of \citet{wuyts08} (hereafter, W08).

The scatter between both the SFR(IR) indicators and SFR({\halpha},{\hbeta}) is larger than the scatter in Figure~\ref{fig:sfrs}(a) by $\sim0.1$\,dex. Incorporating the SPIRE photometry in the CE01 fitting does not change the results, as most of our galaxies are undetected in the SPIRE bands. Whether the increased scatter is due to the limited parameter space of the IR templates or because of ignoring energy balance is an open question that cannot be addressed with the limited data used in this study.

Moreover, we tested the \citet{kennicutt09} empirically-derived coefficients to infer total SFRs by combining observed {\halpha} and IR luminosities. The combined SFRs were systematically lower than our SFR({\halpha},{\hbeta}) and SFR$_{\rm{full-SED}}$ by $\sim0.1$\,dex. The \citet{kennicutt09} sample of nearby galaxies have a higher contribution of evolved populations of stars heating the dust, and therefore, their scaling factor underestimates the total SFR for our star-forming galaxies.

Our sample contains 4 galaxies without {\hbeta} detections, for which we determined $3\sigma$ lower limits on the SFR({\halpha},{\hbeta}). The low SFR$_{\rm{full\,SED}}$ of the {\hbeta}-undetected galaxies compared to the rest of the sample indicates that we are not biased against galaxies in which {\hbeta} is suppressed by dust attenuation. The {\hbeta} line in all four galaxies happened to fall close to a bright skyline, which likely explains their low S/N in {\hbeta}. 

\begin{figure}[tbp]
	\centering
		\includegraphics[width=.48\textwidth]{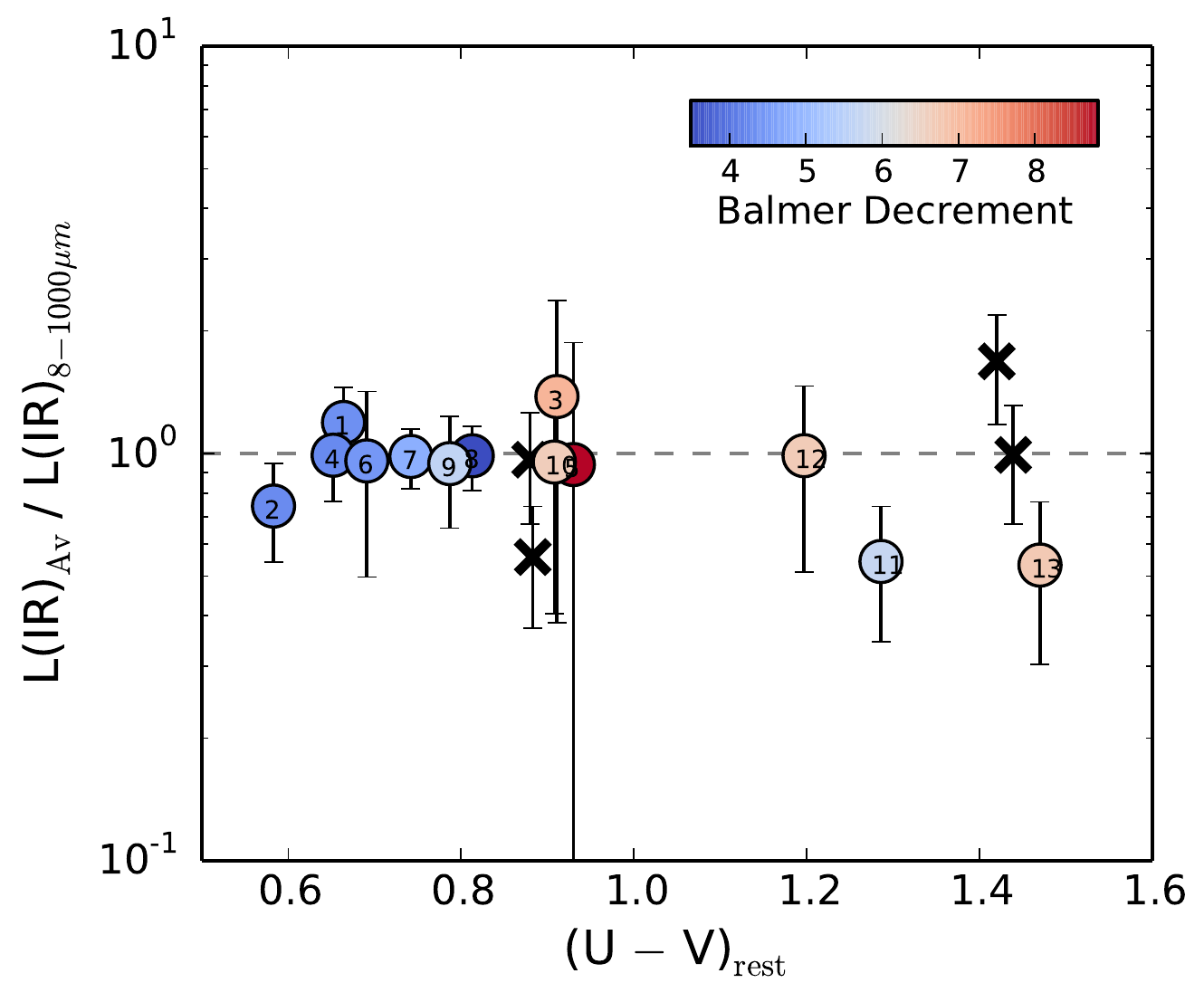}
	\caption{Ratio of the predicted (L(IR)$_{\rm Av}$, based on UV-to-NIR SED) to the measured (L(IR)$_{\rm 8-1000\mu m}$) IR luminosities as a function of rest-frame $U-V$ color. The good agreement between the two estimated IR luminosities suggests that there is not a dominant obscured stellar population that is missing from the UV-to-NIR SED. Galaxies with undetected {\hbeta} are shown with crosses. Numbers correspond to the SEDs in Figures~\ref{fig:seds1},\ref{fig:seds2}.}
	\label{fig:lir}
\end{figure}

\section{Are the IR and UV emission coupled?}

One concern regarding the use of energy-balance SED modeling is that for a galaxy with two distinct stellar populations, with one enshrouded by dust, and hence, bright in IR, and the other population less dusty and brighter in UV, the IR and UV-to-NIR parts of the spectrum will be decoupled. For such cases, an energy-balance method would underestimate the true SFR and L(IR). To address this issue, we calculated L(IR) based on the the UV-to-NIR SED and compared it to the observed $0.3-8\mu$m photometry (using the same stellar population models and fitting assumptions as for the full modeling, Section~\ref{sec:fullsed}). We used the estimated total extinction value at 1600\,\AA~($A_{1600}$) and unobscured SFR(UV) measured from the luminosity at 1600\,~\AA, to calculate the {\em obscured} SFR:
\begin{eqnarray}
 \text{SFR}_{\text{obscured}} = \text{SFR(UV)}~(10^{0.4A_{1600}} - 1.0).
\end{eqnarray} 

Using the \citet{kennicutt98} relation, we converted the obscured SFR to total L(IR) at $8-1000\,\mu$m and compared it to the L(IR) based on our best-fit full SED model, integrated from 8 to 1000\,$\mu$m. The results are shown in Figure~\ref{fig:lir}. For all galaxies in our sample, the UV-to-NIR SED-derived L(IR) is within a factor of $\sim2$ of the integrated full SED L(IR), suggesting that there is not a dominant stellar population that is completely obscured and thus missed from the UV-to-NIR SED.

The results of the UV-to-NIR SED-inferred IR luminosities are sensitive to the assumed SFH and the SED-predicted ages. If instead of a delayed-tau SFH we use a rising SFH of the form ${\rm SFR(}t{\rm )}\propto e^{t/\tau}$, the best-fit SED model predicts a high amount of dust obscuration for our most massive galaxies. Because of the high dust obscuration inferred from the stellar SED, L(IR) and SFR are overpredicted by a factor of $\gtrsim 2$.
Additionally, as found in previous studies \citep[e.g.,][]{wuyts11a}, without a minimum age of 50\,Myr the best-fit UV-to-NIR SEDs result in unrealistically young ages and high dust values for a subset of our galaxies.

The L(IR) comparison test is similar to comparing the UV-to-NIR and full SED SFRs. With a physically reasonable age range and the delayed-tau SFHs, the UV-to-NIR SED and full SED SFRs are in good agreement ($\rho_{\rm s}=0.90$, $3.0\sigma$ away from a zero correlation). Both tests illustrate that the SFR derived from the stellar emission alone is not biased compared to the UV-to-FIR SED-based SFR for $z\sim2$ galaxies with ${\rm M_*}\sim10^{10}-10^{11}\,\msun$ and ${\rm SFR}\sim30-210\,\textrm{M}_\odot\,\mathrm{yr^{-1}}$.

\section{Discussion and Implications}

UV-inferred SFRs are susceptible to missing or underestimating the SFR of galaxies for which the bulk of the star formation is obscured and the UV slope is decoupled from the total obscuration \citep{reddy10,casey14}. Although for a given column of dust {\halpha} and {\hbeta} are less attenuated than the UV, HII regions may also be optically thick at these wavelengths. We show in this paper that for galaxies with ${\rm SFRs}\sim30-250\,\msun$\,yr$^{-1}$ the {\halpha} luminosity, once properly corrected for dust attenuation using the Balmer decrement, do not underestimate the SFR. The same result holds for UV-to-NIR SED inferred SFRs, assuming delayed-tau SFHs. This conclusion does not necessarily hold at higher SFRs (i.e., submillimeter galaxies) owing to highly obscured nebular and/or stellar regions. Such IR-luminous galaxies are likely missed in our study due to their low number density or highly obscured emission lines. To test the latter, we checked the MOSDEF target sample for galaxies with IR detections and undetected {\halpha} lines, and found two. Due to uncertain redshifts and noisy spectra we did not include them in this study. However, deeper and/or wider wavelength range spectra are needed to confirm these systems.

There are additional caveats to our analysis which need to be explored in more detail in future studies. First, both the SFR({\halpha},{\hbeta}) and the SFR$_{\rm{full\,SED}}$ suffer from systematic uncertainties in the assumed IMF, metallicity, and SFH. 
Second, the SFR({\halpha},{\hbeta}) is affected by the choice of the assumed dust attenuation curve (Section~\ref{sec:sfrs}), and the SFR$_{\rm{full\,SED}}$ by the assumed dust emission models. Once we have the complete MOSDEF dataset in hand, we will build upon the results of this study by stacking {\em Spitzer} and {\em Herschel} measurements for our full sample.

~\\
I.S. thanks Mostafa Khezri and Ali Khostovan for useful discussions and feedback on the manuscript. The authors thank Mark Dickinson for providing part of the IR data. I.S. and N.A.R are supported by a National Science Foundation Graduate Research Fellowship DGE-1326120 and an Alfred P. Sloan Research Fellowship, respectively. The MOSDEF survey is funded by NSF AAG collaborative grants AST-1312780, 1312547, 1312764, and 1313171 and archival grant AR-13907, provided by NASA through a grant from the Space Telescope Science Institute.

\end{document}